\documentclass[12pt,prb,aps,notitlepage]{revtex4-1}
\usepackage{amsmath}           		          	
\usepackage{graphicx,epstopdf}					
\usepackage{amssymb}
\usepackage{fullpage}
\usepackage{color}
\usepackage{esint}
\pdfoutput = 1 

\allowdisplaybreaks

\begin{document}
\title{Reply to ``Comment on Nuclear Fusion 66, 016012 (2026) by Richard Fitzpatrick, {\em A Simple Model of Current Ramp-Up and Ramp-Down in Tokamaks}" by A.H.~Boozer }
\author{Richard Fitzpatrick\,\footnote{rfitzp@utexas.edu}}
\affiliation{Institute for Fusion Studies, Department of Physics, University of Texas at Austin, Austin TX 78712}
\begin{abstract}
This report is a follow up to my paper ``A simple model of current ramp-up and ramp-down in tokamaks''  [Nucl.\ Fusion {\bf 66},  016012 (2026)] in the light of comments
on the paper recently made by Dr.~A.H.~Boozer (arXiv:2601.05977). 
\end{abstract}
\maketitle

My recent paper ``A simple model of current ramp-up and ramp-down in tokamaks" \cite{rf} was inspired by a report authored by Dr.~A.H.~Boozer, entitled
``Absolute constraints on the magnetic field evolution in tokamak power plants",  that was published on the arXiv.org repository on July 7, 2025.\cite{boozer}.
In this report, Dr.~Boozer expresses skepticism that the ITER and SPARC design teams have fully taken into account the constraints on tokamak operation
that are imposed by Faraday's law, in particular with respect to the ramp-up and ramp-down of the plasma toroidal current. In Eqs.~(31)-(32) of the report, Dr.~Boozer
estimates that, in the SPARC tokamak, the timescale for the free---that is,  when the primary solenoid flux is held constant---decay of the plasma current from flattop conditions 
is 1157 seconds. In the second paragraph before giving this estimate, Dr.~Boozer mentions that the SPARC design envisages ramping up the
plasma current in 7 seconds, and ramping it down in 12 seconds, and complains that the SPARC team have not fully explained how this is possible.  The implication of these widely differing timescales seems inescapable. If
the timescale for the free decay of the current in SPARC really is 1157 seconds then any attempt to ramp-up the current on a timescale that is a factor of 100
times shorter than this would generate a hollow current profile that is highly unstable to kink modes, which would inevitably trigger a disruption.
Likewise, any attempt to ramp-down the current on a similar timescale would generate a strongly peaked current profile that is highly unstable to
tearing modes, which would again inevitably lead to a disruption. In an earlier paper,\cite{boozer1} Dr.~Boozer seems to imply that an attempt to
ramp-down the current in ITER on a timescale significantly faster than 1000 seconds would lead to a highly peaked tearing-unstable current
profile. However, as noted in the paper, the ITER design envisages ramping down the current in 60 seconds. 

In fact, Dr.~Boozer raises (or appears to raise) a very good point. Ramping the current up and down in a tokamak involves the diffusion of
magnetic flux through the plasma. If the current ramp is substantial then the flux has to diffuse the whole way across the plasma. 
All plasma physics textbooks (including my own\cite{book}) state that the timescale for such flux diffusion is
\begin{equation}
\tau_R = \frac{\mu_0\,a^2}{\eta},
\end{equation}
where $a$ is the plasma minor radius, and $\eta$ the mean electrical resistivity. However, a calculation of this timescale leads to absurdly large
results. For example, a calculation of the  timescale for JET (assuming a mean electron temperature of 7 keV) yields 400 seconds. Despite this, all JET discharges were able to ramp up the
current, achieve a flattop for a significant fraction of the duration of the discharge, and ramp down the current again, in less than 40 seconds.  Obviously,
there is something seriously wrong with the above estimate. (In fact, the referees of my paper, who were clearly very knowledgeable
physicists, and made many valuable suggestions for the  improvement of the paper, were surprised  by how badly the $\tau_R$ estimate misses the mark.)

My paper  constructs a simple model for the current ramp-up and ramp-down process in an ohmically heated tokamak, with the ultimate aim of determining the minimum
safe current  ramp time. The ingredients of the model are uncontroversial. In addition to Maxwell's equations, the model employs a  plasma Ohm's law that relates the
induced plasma current density to the inductive electric field. The Ohm's law involves the plasma resistivity, which is a strong function of the electron temperature.
In order to determine the electron temperature profile, it is necessary for the model to incorporate a temperature evolution equation. The only heating source in an
ohmic plasma is Joule heating, whose form is well known. However, we must also model the transport of electron energy. The transport of
electron energy in tokamak plasma has been the subject of extensive research for the last 50 years. There is overwhelming evidence that
such transport is diffusive in nature. In other words, the transport can be determined by specifying a diffusivity profile, $\chi(r)$. 

Empirical
scaling laws lead to the conclusion that $\chi \sim 1\,{\rm m^2/s}$ in next-generation tokamaks. Thus, we can make an educated
guess as to the average diffusivity. The normalized diffusivity profile, $\hat{\chi}(r)=\chi(r)/\chi_{\rm av}$, is harder to guess. In my analysis, the solution of
the electron transport equation reduces to a nonlinear eigenvalue problem, where the eigenvalue, $\lambda$, is a function of the
normalized $\hat{\chi}(r)$ profile. If $\hat{\chi}(r)$ is uniform then $\lambda=7.01$. If $\hat{\chi}(r)$ increases outward in $r$ (as is likely
to be the case) then $\lambda$ is less than this. However, the variation of $\lambda$ with the $\hat{\chi}(r)$ profile is fairly weak.
In my paper, the numerical estimates are made for a uniform $\hat{\chi}(r)$ profile. However, all of the relevant formulae in the paper
are expressed as functions of $\lambda$. Thus, an interested reader who had a better model for the $\hat{\chi}(r)$ profile
could employ my eigenvalue equation to calculate an improved estimate for $\lambda$, and then use this estimate in my formulae. 

Once the electron transport equation has been solved, my system of equations is characterized by one unknown, which is the inductive
electric field at the plasma boundary, $E_\phi$. Faraday's law gives
\begin{equation}
2\pi\,R_0\,E_\phi =- \frac{d(\psi_p+\psi_s)}{dt},
\end{equation}
where $R_0$ is the plasma major radius, $\psi_p$ the poloidal magnetic flux generated outside the plasma by the plasma current, and $\psi_s$ the magnetic
flux generated by the central solenoid. However, all modern tokamaks are equipped with sophisticated feedback control systems that effectively solve the previous
equation in order to figure out what rate of change of the central solenoid flux is needed to generate a given $E_\phi$ (or, to be more exact, a given
plasma current). In other words, $E_\phi(t)$ is effectively programmable. This being the case, we can work directly in terms of $E_\phi(t)$, 
without explicitly involving $\psi_p$. Of course, this makes perfect sense, because the plasma current profile is determined by
the diffusion of magnetic flux through the plasma itself, rather than through the surrounding vacuum.

A current ramp-up in a tokamak plasma is conventionally accompanied by a simultaneous increase in the plasma minor radius, and vice versa 
for a current ramp-down. My analysis shows that it is possible to effect a current ramp-up or ramp-down in which the safety factor
profile is held constant provided that the plasma minor radius and $E_\phi(t)$ are adjusted in a judicial fashion, and the ramp-up/ramp-down rate is
smaller than a critical rate.
 Obviously, if the initial safety-factor profile is stable to kink and tearing modes then the plasma
will remain stable to such modes throughout the current ramp. I find that the minimum safe ramp time is actually much less than
$\tau_R$ for a number of reasons. First, because $\lambda\sim 7$, rather than $\sim 1$. Second, because the  plasma  during the current ramp 
is much colder than the flattop plasma, because the latter plasma is subject to strong auxiliary heating, whereas the former
 is only subject to Joule heating, and is, therefore, not as hot as the flattop plasma. Third, because the plasma
minor radius is significantly less than the flattop minor radius during most of the ramp, so magnetic flux has a smaller distance to
diffuse. In fact, my estimates for the minimum safe ramp time are entirely consistent with previous experimental results, and  also
with the SPARC and ITER designs. 

The final takeaway from my paper is that Faraday's law does not impose any constraints---absolute, or otherwise---on tokamak
operation that are incompatible with the present SPARC and ITER designs. This, of course, is hardly a surprising result, because both machines were designed with the aid of computer 
simulations that effectively perform far more sophisticated versions of the calculation described in my paper. The only merit of my calculation is that it is sufficiently simple that an
interested reader can easily follow all of the steps, and so verify the conclusions.

Dr.~Boozer has suggested\,\cite{boozer2} that I have misrepresented his views, whose evolving, and somewhat nebulous, form is documented in the four subsequent revisions of the initial version of the report in question.\cite{boozer} 
However, not only did Dr.~Boozer post the initial version of his report publicly, he also distributed it among members of a committee (that included both myself and Dr.~Boozer) that had been tasked
by the U.S.\ Department of Energy, Office of Fusion Energy Sciences to advise them on future directions for the U.S.\ plasma fusion theory program, and recommended it to the members of the committee for their urgent perusal. 
The avowed purpose of Dr.~Boozer in distributing his report was to initiate a debate within the fusion community on the issues raised in the report. In this respect, he succeeded. 
I think that my interpretation of Dr.~Boozer's initial version of his report is a reasonable one. Dr.~Boozer has also claimed that my paper makes fundamental errors in physics. Given
that the paper has now been published in a refereed journal, after a rigorous review process, I am happy to leave it to the fusion plasma physics community to determine whether or not this is the case. I believe that I have answered the main criticisms of
the paper in this report.

\end{document}